\documentclass{article}
\pdfoutput=1
\usepackage{arxiv}

\usepackage[utf8]{inputenc}
\usepackage{amsmath}
\usepackage{adjustbox}

\DeclareMathOperator*{\argmax}{argmax}

\usepackage[sort&compress]{natbib}

\newcommand{\epochs}{50}

\setcitestyle{super,square,compress,citesep={,}, notesep={;}}
\makeatletter
\def\NAT@def@citea{\def\@citea{\NAT@separator}}
\makeatother

\usepackage{natmove, graphicx, wrapfig, caption, subcaption, float, url}

\title{Investigating Active Learning and Meta-Learning for Iterative Peptide Design}

\author{
    Rainier Barrett\\
    Department of Chemical Engineering\\
    University of Rochester
    \And
    Andrew D White\thanks{andrew.white@rochester.edu}\\
    Department of Chemical Engineering\\
    University of  Rochester\\
}

\date{\today}

\begin{document}

\maketitle





\begin{abstract}
Often the development of novel functional peptides is not amenable to high throughput or purely computational screening methods. Peptides must be synthesized one at a time in a process that does not generate large amounts of data. One way this method can be improved is by ensuring that each experiment provides the best improvement in both peptide properties and predictive modeling accuracy. Here, we study the effectiveness of active learning, optimizing experiment order, and meta-learning, transferring knowledge between contexts, to reduce the number of experiments necessary to build a predictive model. We present a multi-task benchmark database of peptides designed to advance these methods for experimental design. Each task is binary classification of peptides represented as a sequence string. We find neither active learning method tested to be better than random choice. The meta-learning method Reptile was found to improve average accuracy across datasets. Combining meta-learning with active learning offers inconsistent benefits.
\end{abstract}



\section{Introduction}

Although great strides have been made in predictive computational modeling where large datasets exist, the use of computational modeling where data is scarce and/or expensive has been limited. Data expense and scarcity is the norm in materials design\cite{Hutchinson2017}. Computer-aided design often relies on physics-based predictive methods \cite{Shi2017,Afzal2019}, which require no data, but cannot predict complex properties like activity of a drug or refractive index of a thin film. Even if physics-based modeling is used, there is no clear mechanism to improve predictions as data is gathered in the course of testing. Thus, human intuition is often the state-of-the-art for choosing which new peptides to test when there are small amounts of data.

 Peptides are a popular target for biomaterials design.\cite{Krishna2010PeptideBiopolymerReview,Mart2006PeptideResponsiveBiomaterials,Zaccaria2018PolymerPeptideBiomaterials,Wang2017PeptideBiosensors,DeMel2008BiomaterialsReview,Hamley2017BiomaterialsReview} Unlike most synthetic polymers, the amino acid sequence of a desired peptide can be controlled at the monomer level to form specific sequences.\cite{Chan1999FMOCPeptideSynthesis} Being biologically-derived polymers, peptides are generally bio-compatible. They are relatively easy to synthesize, and some undergo self-assembly to form ordered structure spontaneously.\cite{Cui2010PeptideSelfAssembly,Chetia2020PeptideOrganogels} Creating or coating an object with functional peptides is an appealing way to create bio-compatible materials with various applications in medicine,\cite{Hosoyama2019PeptideTissueRepair,TabatabaeiMirakabad2019PeptideDendrimerMaterials} drug delivery,\cite{Shah2018PeptideDrugDelivery,Varanko2020PeptideDrugDeliveryReview} and more.\cite{Sanghvi2005ConductorBindingPeptides,Chetia2020PeptideOrganogels,Zong2017PeptideNanoparticles,Zaccaria2018PolymerPeptideBiomaterials,Wang2017PeptideBiosensors} 

Here we apply active learning to binary classification of peptides across a variety of binary tasks like predicting solubility or activity against bacteria. We examine two standard active learning methods: query by committee (QBC)\cite{FreundQBC1997} and uncertainty minimization\cite{lewis1994sequential} with supervised learning of a deep convolutional neural network. We also examine meta-learning to see if there is a benefit in transferring knowledge from one task to another. We only allow our active learners to choose from the dataset of labeled peptides (rather than generate new sequences), so that we can accurately assess the selection during training. However, these methods are ultimately evaluated not based on the peptide chosen, but the accuracy of the resulting trained model. The rationale is that there are many competing design constraints in peptide design (e.g., synthetic feasibility, cost, bioavailability, etc) and thus it is better to have an accurate model than a finite set of examples proposed to be active. 

Computer-aided design of bio-active peptides is an established field of research with a variety of approaches, spanning from quantitative structure-activity relationship (QSAR) modeling to machine learning. For example, Franco et al.\cite{Franco2018GeneticPeptides} used a genetic algorithm to optimize an antimicrobial peptide derived from the guava plant, and Hancock et al.\cite{Hancock2018BiofilmQSPR} used a QSAR model to find anti-biofilm peptides. For further information on peptide design principles and modeling efforts, see \citet{Nunez2019PeptideDesignReview}, \citet{Schneider2012AMPDesignReivew} and \citet{Cronin2019ChemicalSpaceReview}.
The goal of this work is not to compete with these methods, but to assess active learning and meta-learning as potential ways to improve iterative discovery of peptides in this setting. 

Active learning has a history as an extension to design of experiments, which is about choosing the optimal experiments to do with limited resources. 
Our concern is a sequence of experiments where the results of the previous experiments influence our decision of the next, whereas optimal design of experiments is about choosing the best experiment prior to beginning, and assumes a linear model. Active learning is this process of choosing the next experiment optimally\cite{Settles2010}. It is sometimes called optimal experimental design\cite{Liepe2013}, targeted experiment design\cite{Vanlier2012}, sequential design of hypotheses\cite{ZACKS1996151}, optimal learning\cite{Gopakumar2018}, and artificial intelligence scientific discovery\cite{Buchanan1968} depending on the goal and problem context.

Within this framework, there are a variety of approaches depending on the form of the task model and utility function. If the task model is probabilistic, like above, utility functions which maximize information gain\cite{Li2013a,Liepe2013}, reduce model uncertainty\cite{5509293}, maximize expected model change\cite{Settles2010}, or reduce model variance\cite{Mackay1992} can be chosen. Within variance reduction methods, there are so-called A-optimal\cite{Mackay1992}, D-optimal\cite{Mackay1992,Chaloner1993}, and E-optimal\cite{Flaherty2005} approaches which minimize the covariance matrix according to different assumptions. If the task model is non-parametric, Bayesian approaches are well- suited\cite{Vanlier2012}. One still has a choice of utility function (also known as acquisition function) and can, for example, maximize the expected model improvement at each experiment\cite{Chaloner1993,kapoor2007active}. Bayesian approaches can work with recent deep learning methods through Bayesian convolutional neural networks \cite{Gal2017} or through the connection between dropout and neural network uncertainty\cite{Gal2015}.

If the task model is not stochastic but there is flexibility in parameter choice or multiple models to choose from, then there are a variety of pooled or consensus active learning approaches. Query By Committee (QBC) is an active learning approach that maintains a committee of models and chooses the next experiment based on where the committee disagrees most\cite{seung1992query}. This ``disagreement'' is quantified using one of the various utility functions described above, for each committee member, and combining the results in some way, e.g. using summation or taking the mean. There are also active learning pool methods for specific model types. For example, k-nearest neighbor\cite{wei2015submodularity}, logistic regression\cite{hoi2006batch,guo2007discriminative}, and linear regression with noise\cite{yu2006active}. One can also treat the choice of model as a probability distribution and then the pool of models can be viewed as a stochastic or Bayesian model to use any of the previous utility functions.

There are active learning methods that are independent of the task model and instead find a characteristic set of data\cite{cohn1996active}. This can be done by clustering the data, optimizing a function which describes local variance\cite{Yang2015}, finding regions of high uncertainty\cite{cohn1994improving}, or by finding a  characteristic subset of data called a ``coreset'' via submodular function optimization\cite{Sener2017,wei2015submodularity}. These methods are closely related to semi-supervised learning, which tries to use unlabeled data to improve a model when labeled data is sparse\cite{corduneanu2002information,szummer2003information}.

Another closely related topic to active learning is Bayesian optimization or global optimization of black box functions. There the goal is to optimize a function while evaluating it a minimum number of times. To connect this to active learning described above, view the ``expensive function'' as the experiment. A surrogate stochastic model is constructed, often through bootstraping or non-parametric models, and that surrogate model is used within an active learning framework to reduce the number of the function evaluations\cite{jones1998efficient}. Then active learning is applied so that each function evaluation improves the maximum function value and/or understanding of the model. This method can be equivalent to the variance reduction techniques discussed above if the same utility functions are chosen\cite{chaloner1995bayesian}. More sophisticated active learning methods can be used with the surrogate model, including Gaussian process regression and complex portfolios of acquisition (utility) functions\cite{hoffman2011portfolio,shahriari2014an}. A common critique of Bayesian optimization is that it typically scales between $O(n^2)$ to $O(n^3)$, depending on approximations made, and struggles with high-dimensional surrogate models. However, this is not applicable here, since our goal is learning in systems with dozens of experiments, not thousands. A recent overview on the application of Bayesian optimization to materials design can be found in Frazier and Wang.\cite{frazier2016bayesian}

An early example of active learning in chemistry can be found with van de Walle et al.\cite{van2002automating}, where a phase diagram was explored using variance minimization active learning on cluster expansions. Active learning works well in general with choosing cluster expansions via variance minimization\cite{Mueller2010,PhysRevB.83.224111}. More recently, Lookman et al.\cite{Lookman2017} explored elastic properties with ab initio calculations using a Bayesian optimization technique (Efficient Global Optimization) to optimize the ratio of three metals. The authors applied the same method to design new piezoelectrics\cite{yuan2018accelerated} with a four dimensional design space. Gopakumar et al.\cite{Gopakumar2018} also showed how active learning methods that balance exploration and exploitation can work well on 2 and 3 dimensional systems. This active learning approach to finding compositions with Bayesian optimization is quickly gaining popularity in the informatics community for low-dimensional systems\cite{Fukazawa2019a,Wen2019a,Rickman2019c}.

Kim et al.\cite{Kim2019} explored active learning methods to find polymers with a specific glass transition temperature. This is a high-dimensional system because the polymers are represented with a variety of descriptors. It was made tractable by treating a list of 731 possible polymers as known a priori. At each step the model evaluates all 731 possible polymers. An earlier example using a fixed molecule set can also be found in Warmuth et al.\cite{warmuth2002active} where candidate drug molecules were selected from a vendor catalogue with a variance minimization active learning algorithm.

Recently \citet{Tallorin2018DiscoveringLearning} explored the use of Bayesian optimization for de novo design of peptide substrates. Their work is similar in that both this work and theirs used a sequence model with the goal of minimizing the number of experiments required to train a model. The differences are that they were modeling with regression, allowed for complete choice in sequence space, did not have a goal of creating accurate models, and did not use a deep learning model but a Na\"ive Bayes classifier. Their work is an experimentally validated demonstration that intelligently designing experiments with predictive models can reduce the required number peptides that need to be synthesized and tested. 

Another topic explored in this work is meta-learning. Meta-learning is a technique for improving few-shot learning across multiple tasks. The goal, in our nomenclature above, is to make the task model depend on hyperparameters $\xi$ that are trained to work well across multiple tasks. Then on a new task, using $\hat{\xi}$, few new examples are required to improve performance. Active learning and meta-learning are connected because both are concerned with maximizing the value of data. As the goal is to minimize the number of experiments required for new systems, we find it natural to consider this method on our dataset. Examples of meta-learning for accelerating task models can be found in transfer learning\cite{Pan2010}, few-shot learning\cite{Altae-tran2017, Snell2017, Vinyals2016}, and automated machine learning\cite{quanming2018taking}. One application that has recently connected active learning and meta-learning is in model free reinforcement learning\cite{Duan2017,Gupta2018}. \citet{Pang} and \citet{Fang2015} have also explored the connection between active learning and meta-reinforcement learning.

\section{Methods}

\begin{table*}[ht]
    \small
    \centering

    \makebox[\textwidth][c]{
    \begin{tabular}{|l|c|r|}
    \hline
    Positive Dataset & Size & Negative Datasets\\\hline
    antibacterial & 2079 & shp2$^1$, tula2, insoluble$^2$, antifungal$^3$, antiHIV, anticancer$^4$, scrambled\\
    anticancer & 183 & shp2, tula2, insoluble, antifungal, antiHIV, antiparasital, antibacterial$^4$, scrambled\\
    antifungal & 891 & shp2, tula2, insoluble, antiHIV, anticancer, antibacterial, scrambled\\
    antiHIV & 87 & shp2, tula2, insoluble, antifungal, anticancer, antiparasital, antibacterial, scrambled\\
    antiMRSA &  119 & shp2, tula2, insoluble, antiHIV, anticancer, antiparasital, scrambled\\
    antiparasital & 90 & shp2, tula2, insoluble, antiHIV, anticancer, scrambled\\
    antiviral & 150 & shp2, tula2, insoluble, antifungal, anticancer, antiparasital, antibacterial, scrambled\\
    hemolytic & 253 & shp2, tula2, insoluble, human$^5$, scrambled\\
    soluble & 7028 & insoluble$^6$\\
    shp2 & 120 & scrambled$^7$\\
    tula2 & 65 & scrambled$^7$\\
    human & 2880 & insoluble, hemolytic, scrambled\\
    \hline
    \end{tabular}
    }
        \vspace{0.2cm}
        \caption{The positive and negative examples chosen for training the classifiers. Negative datasets were sampled to be the same size as the positive datasets. $^1$ It is exceedingly rare to find antibacterial peptides, so it is assumed that the SHP-2/Tula-2 binding peptides are not good examples of antibacterial peptides. $^2$ Insoluble peptides cannot be successful antibacterial peptides. $^3$ Antifungal and antibacterial activity are different mechanisms, so it is unlikely that a given peptide is both antifungal and antibacterial. $^4$ It is known that antimicrobial and anti-cancer peptides often have a similar method of action, and there is significant overlap between the two datasets. Including these datasets in one another's negative example sets might be expected to reduce overall model accuracy. This conservative choice of dataset still resulted in accuracy near baseline on these two tasks in the context of meta-learning and active learning.} $^5$ it is assumed that fragments of proteins found on surfaces of proteins found in humans are not hemolytic (kill red blood cells). $^6$ No scrambled scrambled dataset is necessary because there are known negative examples. $^7$ classifying SHP-2 and Tula-2 is in the context of fixed-length peptides so only scrambled peptides with the same length are used for classification. 
    \label{tab:negatives}
\end{table*}

Functional peptide datasets were prepared by mapping each amino acid to a one-hot encoded vector of dimension $[N\times 20]$, where $N$ is the length of the peptide. Five past published databases were used to generate training datasets in this work: the Antimicrobial Peptide Database (APD)\cite{apd3}, the collection of protein surface fragments from White {\textit et al.},\cite{WhiteHumanFragments2012} the PROSO II database,\cite{Smialowski2012PROSOPrediction} library hits with activity against TULA-2 protein\cite{Cheng2010}, and library hits with activity against SHP-2\cite{Sweeney2005SHP}. The APD contains peptides flagged with a variety of activities such as antibacterial, antifungal, antiviral, anticancer, and hemolytic activities. The PROSO II databse contains peptides and proteins categorized as soluble or insoluble. The TULA-2 and SHP-2 libraries contain fixed-width peptides optimized for binding to a specific target. Eight datasets were chosen from the APD: 1. ``antibacterial,'' 2. ``anticancer,'' 3. ``antifungal,'' 4. ``antiHIV,'' 5. ``antiMRSA,'' 6. ``antiparasital,'' 7. ``antiviral,'' and 8. ``hemolytic.'' All sequences above length 200 were excluded. This wide variety of tasks represents a range of modeling goals in peptide research. Table \ref{tab:negatives} shows the negative datasets used for training each classifier. Here, ``human'' refers to the aforementioned collection of human protein surface fragments.

Active learning is typically formulated as an optimization problem\cite{Settles2010}. Consider $N$ observation pairs $x_i, y_i$ of features and labels, respectively, with $i\in [0, 1, \ldots , N]$ indicating the order of observation. Assume that $y_i$ is a class label and $x_i$ is a feature vector of real numbers. We have a \textit{task model}, $P_{\theta_i}(y|x)$, that assigns a probability to each class label for a feature $x$ and is defined by parameters $\theta_i$, which are updated after each new observation. In this work $P_\theta$ is a deep-learning convolutional neural network. $\theta_i$ is updated according to some training procedure after a new $x_i, y_i$ pair is observed. In active learning, we choose $x_{i+1}$ from our fixed dataset of $x,y$ pairs according to
\begin{equation}
\label{eq:al}
x_{i + 1} = \argmax_{x_j} A_{\psi}\left[x_j, P_{\theta_i}(y_j|x_j)\right]
\end{equation}

where $A_{\psi}(\cdot)$ is a functional of the task model and possibly $x$. The $j$ subscript only indicates that for a given $x_j$ we must use the corresponding $y_j$ to evaluate the functional. $A_{\psi}(\cdot)$ can be defined by parameters $\psi$, although it is normally fixed. For example, $A_{\psi}(\cdot)$ could be the most uncertain point, which gives

\begin{equation}
x_{i + 1} = \argmax_{x} \left[1 - P_{\theta_i}(\hat{y}|x)\right]
\end{equation}
where $\hat{y}$ is the most likely class label for $x$ under $P_{\theta_i}$\cite{lewis1994sequential}. $A_{\psi}(\cdot)$ is called the acquisition function or utility function depending on the problem setting\cite{Settles2010}.

Here, the task model is a deep convolutional neural network classifier, necessitating negative training examples as well as positive. One corresponding negative training data set was generated for each positive data set. Two types of negative data were generated: scrambled scrambled data with amino acid distributions identical to the ``soluble'' dataset but length distributions identical to the corresponding positive dataset, and samples from datasets which are expected to have no intersection with positive examples (see Table~\ref{tab:negatives}. These are expected to be rather challenging negative examples, since scrambled peptides likely have many physical properties of the positive examples. Generally, the non-intersecting datasets are also naturally occurring peptides that likely are biologically relevant. To generate the scrambled scrambled negative data set, a number of peptides were generated randomly, with lengths sampled from the same length range as their respective positive set, and residues sampled from the frequency distribution of the soluble dataset. The non-intersecting examples are shown in Table~\ref{tab:negatives} along with rationale in the caption. To balance classes, the negative datasets were sampled down to be the same size as the corresponding set of positive examples. 

Peptide sequences are encoded as one-hot vectors of dimension $N\times20$ (with $N$ the length of the peptide), where each index in the second dimension corresponds to the index of the amino acid in the $n$th position in the alphabet of amino acids:  [A, R, N, D, C, Q, E, G, H, I, L, K, M, F, P, S, T, W, Y, V]. Activity was encoded as a one-hot label vector of length 2, where  $[1,0]$ indicates a positive label and $[0,1]$ indicates a negative label. 

\begin{figure}
    \centering
    \includegraphics[width=0.5\linewidth]{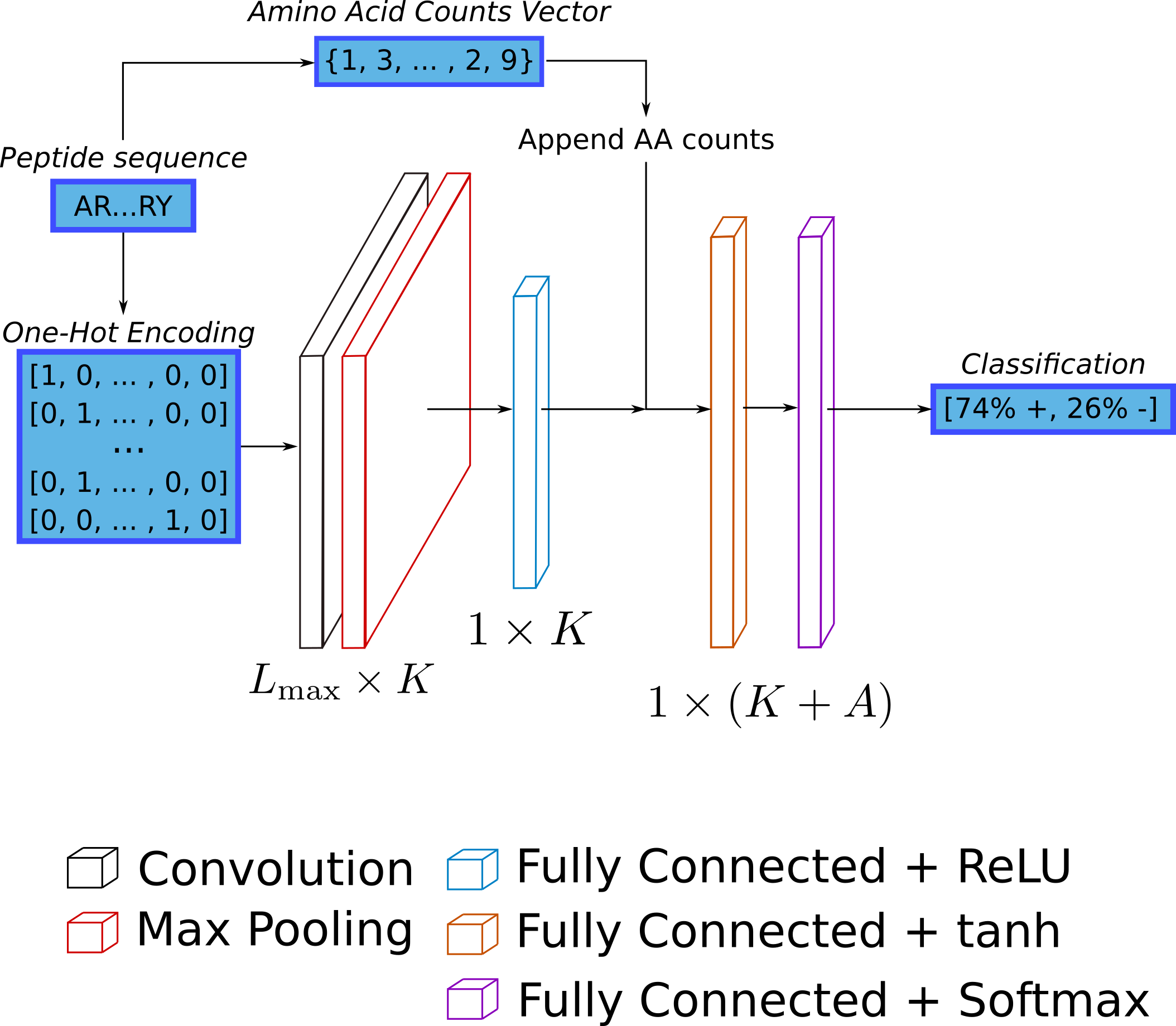}
    \caption{Neural network structure for active learning. Here, $L_{\text{max}}$ is the maximum width of a peptide in the dataset (though a convolution can use any length), $K$ is the number of motif classets, and $A$ is the length of the amino acid alphabet. Peptides are first translated to a one-hot encoding ($L_{\text{max}} \times A$) and a vector of normalized amino acid counts ($1\times N$). The output of the max pool layer is passed through one fully connected layer with ReLU activation, then amino acid counts are appended to the output. This is then passed into two more fully connected layers with a final output dimension of $2$, for positive and negative class labels. Labels below neural network layers indicate the dimensionality of the data as it passes through the layer.}
    \label{fig:CNN_diagram}
\end{figure}

The task model is a convolutional neural network whose structure was partially based on that of a model from our past work in peptide modeling.\cite{Barrett2018} The model structure is shown schematically in Figure~\ref{fig:CNN_diagram}. The first layer of the neural network is a convolutional layer with a weight matrix of dimension $[W \times A \times K]$, where $W$ and $K$ conceptually represent peptide ``motif widths'' and number of ``motif classes,'' respectively, and $A$ is the length of the amino acid alphabet considered ($20$ for the naturally-occurring peptides used here). The next layer is a mean pool layer, which captures which ``motif class'' is most likely in the input peptide by pooling across the peptide's length dimension, leaving a $1\times K$ vector.

The output of the max pool layer is concatenated with the relative frequency of each amino acid in the input peptide ($1\times 20$ vector) and standardized ($0$\--$1$) sequence length. This is then fed to three fully-connected (FC) layers with ReLU activation function and then to one FC layer with softmax activation for classification, ensuring the outgoing vector of size 2 adds up to 1, since this vector is meant to represent the likelihood of assignment to the positive or negative classes. The final output is compared with the true label vector (classification) or the true activity (regression), and loss is calculated as the cross-entropy between the two. The minimization algorithm used during training was TensorFlow's\cite{tensorflow2015whitepaper} Adam optimizer with parameters recommended in \citet{KingmaAdam2014} and a learning rate of 0.001. This architecture is relatively simple compared to state-of-the art deep leanring for sequence prediction tasks of peptides. Recently, \citet{VeltriDeepLearningPeptides2018} developed a deep learning model for predicting antimicrobial activity. As our goal is to explore active and meta-learning, we use a simpler architecture. Namely, we do not use embeddings, and we use smaller and fewer filters, and fully-connected layers after features instead of a recurrent neural network.

The method of uncertainty minimization is designed to favor exploration to maximize information gained per training iteration of a model. This is achieved by choosing a new training point based on some measure of the uncertainty of the machine learning model used. In this sense, it is well-aligned with an eventual goal of automated experiment selection, because it can minimize the number of necessary experiments to characterize a property space well. This, in turn, would lead to a reduction in operating costs and time spent performing experiments. 

For uncertainty minimization training, the model described previously was used, with  $W$ and $K$ both chosen to be 6. Before training, model weights are randomly initialized. The model uncertainty is calculated as the variance of the output vector of the neural network for each peptide. One peptide is then sampled, with probabilities of selection for each peptide weighted by their respective variances under the current model parameters, i.e. $p(1-p)$. This is different than Equation~\ref{eq:al}, where $\argmax$ is used instead of sampling. The chosen peptide is then used to train the neural network for one training iteration, then an additional 16 training iterations are performed with batch size 16, sampling uniformly from all previously-observed peptides for training input. Thus, in the first iteration the first point is used for \epochs{} training steps, then in the second, the first and second points are used an expected value of 12.5 times each, etc. This process is repeated for \epochs{} training epochs for one training run. To gather statistics, 30 training runs of \epochs{} epochs each were performed for each dataset. We also investigated model performance after 100 training runs of 10 epochs, to evaluate the few-shot learning capability of the various combinations of active and meta-learning methods used here.

The QBC method employed in this work uses the same approach as uncertainty minimization, but instead of a single task model that is trained and used for selection, a committee of 10 models is used. The 9 additional models are all structured in the same way as the model used in uncertainty minimization, but use different hyperparameters. They differ in the dimensions of the weights matrix used in the convolution layer, having all combinations of $3 \leq w \leq 5$ and $3 \leq k \leq 5$ along with the $W = 6,$ $K = 6$ model used in uncertainty minimization. In QBC, input data is passed to each committee member (task model), and each one produces a prediction. The average variance among all models is used as sampling weight for selecting a new training point, and training is performed in the same way as for uncertainty minimization, with the same number of epochs and training runs.

To compare these active learning methods against a base case, we use two control training methods with the same model as used for the uncertainty minimization method. The first control is a ``baseline'' Adam training, where the model trains on all peptides for 5,000 steps with a batch size of 32. The second is a more direct comparison to the active learning methods called ``random'' where peptides are chosen randomly and the model is trained in the same way as in the active learning methods (batch size 16, 16 iterations, \epochs{} epochs). 

The meta-learning method used in this work is Reptile from \citet{Nichol2018}. It is related to the the work by \citet{Finn2017}, called model agnostic meta-learning. In our notation, the goal of meta-learning is to optimize the initial parameters of the task model, $\theta_0$, to work well given a sampled task $\tau$. $P(\tau)$ is taken to be uniform across our datasets. $\theta_0$ is optimized by Adam optimization of a meta-objective function:

\begin{equation}
 \mathrm{E}_{P(\tau)}\left[\mathcal{L}\left[U(\psi, \theta_0, \tau)\right]\right]
\end{equation}

where $\tau$ is the dataset corresponding to a set $(x,y)$, $\psi$ are the parameters defining the active learning method $A_\psi(\cdot)$, and $\theta_0$ is the given initial task model parameters. $\mathcal{L}$ is the usual loss function and $U$ is a stand-in for doing $J$ steps of active learning training with $A_\psi(\cdot)$. The gradient of this meta-objective requires a Hessian, but Reptile approximates this with a Taylor expansion. In this work $J=16$, meaning we train with active learning on 16 peptides each time with a batch size of 16. 2,500 meta-learning iterations were done and then early stopping was used to prevent overfitting. This was done on an 80\% split of the left-out dataset and then results were reported on the 20\% remaining sequences of the left-out dataset.

AUC values and accuracies are reported on withheld data which was 20\% of the dataset size. Training curves are shown in Figures~\ref{fig:umin-results} and \ref{fig:qbc-results}, and exact final values for withheld set accuracy and ROC AUC values are reported in the Supporting Information in Tables S1-S4. The accuracies and AUCs reported here are on the withheld dataset from training. During meta-learning, the location of active/inactive labels (first or second index) was swapped between tasks to prevent over-fitting to active being in one position or another. This gives minimal zero-shot accuracy, but helps illustrate the rate of training, as shown in Figures~\ref{fig:umin-results} and \ref{fig:qbc-results}. The minimized loss function was cross entropy between label probabilities and true labels. Error bars and individual traces are different due to split differences of data and random number initial parameter seeds. 

\section{Results and discussion}

\begin{figure}[ht]
    \centering
    \includegraphics[width=\linewidth]{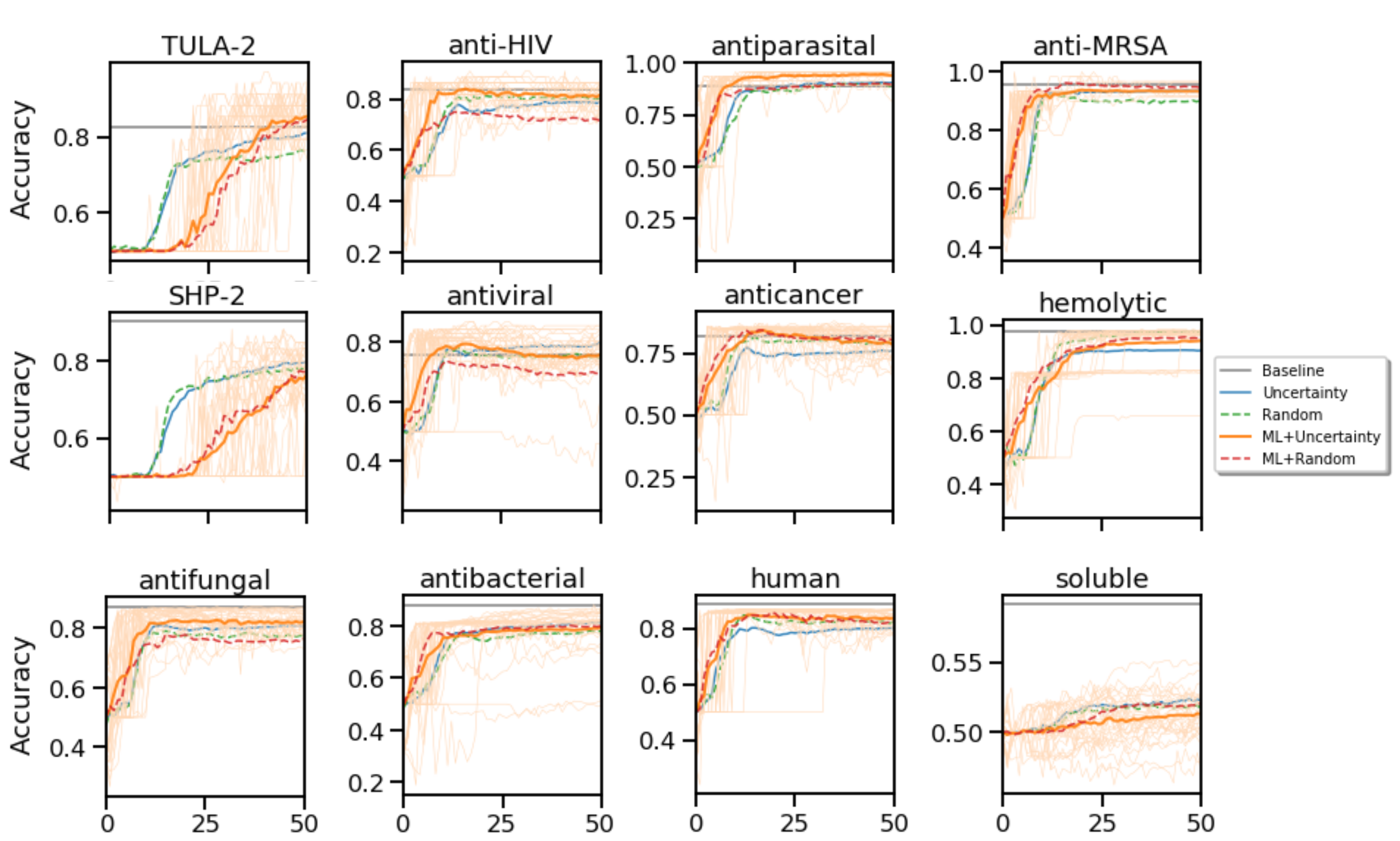}
    \caption{Training curves of uncertainty minimization active learning compared with baseline (gray) trained across all data points and randomly choosing examples. The y-axis is accuracy on withheld data. Light traces are individual 30 runs and the dark trace is median (only one set of traces is shown). Each run has a different train/withheld split and random number generator seeds. Each subplot is a different task, arranged in increasing order of number of training points from left to right, top to bottom.}
    \label{fig:umin-results}
\end{figure}

\begin{figure*}[ht]
    \centering
    \includegraphics[width=\textwidth]{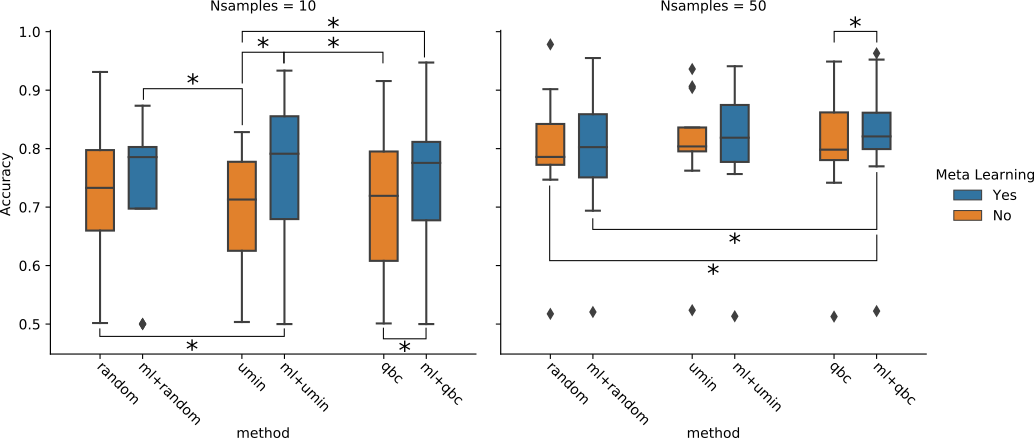}
    \caption{Box-and-whisker plot comparing average accuracy values after 10 and 50 training examples across all twelve datasets for five different methods explored in this work. Asterisks indicate statistically significant difference ($p\leq 0.05$) in mean accuracy, calculated using Wilcoxon's signed ranks test among the twelve average accuracy values compared between two methods. Meta-learning significantly improves few-shot performance over uncertainty minimization or QBC alone when combined with these active learning methods, but only ML+QBC shows significant performance improvement after 50 training examples. This indicates that meta-learning can be a good tool for increasing few-shot learning in settings where data is scarce. 
    See Tables S7 and S8 for tables of all possible p-value pairs.}
    \label{fig:barchart}
\end{figure*}

\begin{figure}
    \centering
    \includegraphics[width=\linewidth]{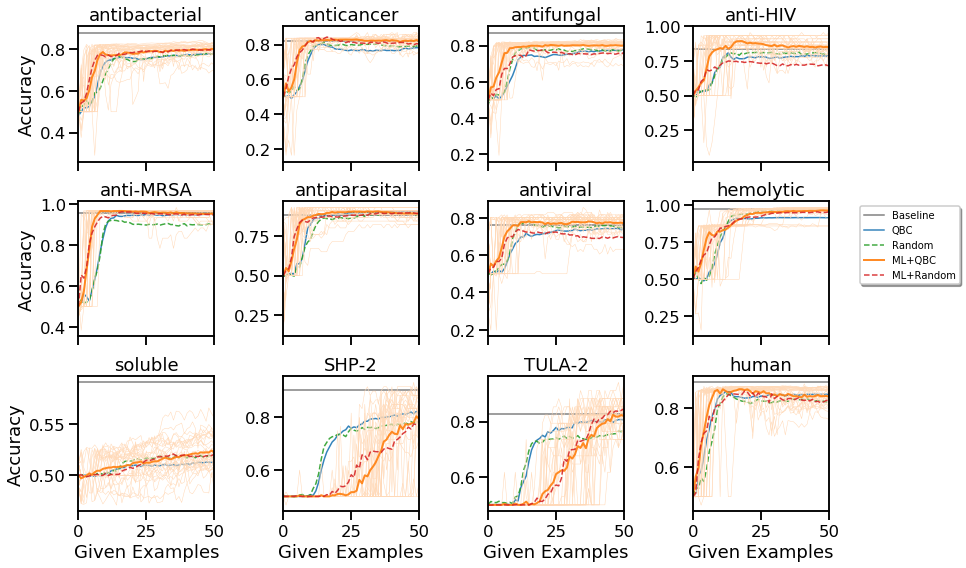}
    \caption{Training curves of uncertainty minimization active learning with and without Reptile meta-learning, compared with baseline (gray) trained across all data points and training with randomly-chosen examples with and without meta-learning. The y-axis is accuracy on withheld data. Light traces are the \epochs{} individual runs and the dark trace is the median. Each run has a different train/withheld split and random number generator seeds. Each subplot is a different task which was withheld during meta-learning. Meta-learning offers inconsistent improvements, while active learning consistently offers no improvements over random, unless paired with meta-learning (though still not consistently).}
    \label{fig:qbc-results}
\end{figure}

Figure~\ref{fig:umin-results} shows the active learning, meta-learning, and baseline results across the twelve datasets for the uncertainty minimization active learning strategy. The baseline results (gray) show that the convolutional neural network provides reasonable results across the range of modeling tasks despite its simple architecture.  The sub-panels are arranged in increasing order of training set size, which were 65, 87, 90, 119, 120, 150, 183, 253, 891, 2079, 2880, 7028, respectively. Overall, training set size does not correlate to learning performance. The solubility task showed the worst overall performance  despite being the largest dataset. Solubility classification is challenging because many of the training examples are folded proteins, requiring long-range sequence correlations to model properly.  The length range of the solubility set is large, ranging from 19 to 198 amino acids. However, the range of the antifungal (2 to 143) and antibacterial (2 to 183) is similar, so this is not likely the source of difficulty. Indeed, solubility prediction is known to be a difficult task, with state-of-the-art models achieving reported accuracies between 0.52-0.77.\cite{Khurana2018DeepSol:Prediction,Chang2013SolubilityAccuracies} Our simple convolutional neural network has an accuracy of 0.59, which is within this range.  

Figure~\ref{fig:umin-results} also shows the comparison of choosing peptides randomly and choosing peptides for which the model has maximum uncertainty (uncertainty minimization). Uncertainty minimization (blue line) is not better than choosing randomly (dashed green) in general. It is sometimes worse and sometimes better in the long term, but for few-shot learning of up to 10 training examples, it is not significantly different from random choice (see Figure~\ref{fig:barchart}). 

To assess the effect of meta-learning on reducing experiment number, it was evaluated both alone and in combination with active learning. Results from combining meta-learning with uncertainty minimization are shown in Figure~\ref{fig:umin-results}. Meta-learning consistently improves initial gains in accuracy, except in the soluble, SHP-2 and TULA-2 tasks. This is likely because these tasks are quite different from the other nine, so feature re-use is less important.  Note that meta-learning results were controlled for label correlation due to dataset overlap by randomly swapping labels (negatives become positive and vice-versa) between meta-training trials. Thus, although many negative datasets overlap -- as do the positive sets in the cases of anti-cancer and antibacterial -- meta-learning is unable to over-fit to the labels of any one set.

Combining uncertainty minimization with meta-learning only sometimes increases accuracy over random choice, but significantly improves few-shot accuracy on average after 10 training examples (see Figure~\ref{fig:barchart}). In some datasets, final accuracy approaches or even exceeds baseline levels with less than 25 examples, whereas the baseline is trained on all data. This demonstrates the advantage of using meta-learning.

Receiver operator characteristic (ROC) curves provide an accounting of the balance between type I and type II error. This is important for peptide activity because, due to the large design space, false positives are more detrimental to a model's usefulness. The area under curve (AUC) of a ROC curve gives a scalar representing the quality of the ROCs. Note that here we enforced balanced classes (same number of positive/negative examples in the training set). This ensures that accuracy and ROC are good measures of performance, while other metrics become necessary in cases with unbalanced training sets. The ROC AUCs are reported in the Supporting Information in Tables S1 and S2 and Figure S1. As observed in Figures~\ref{fig:umin-results} and \ref{fig:barchart}, there is no significant gain to be had in using uncertainty minimization active learning. Also, \epochs{} examples is not enough to match the baseline models without meta-learning. The source of variance in this work is because there are many ways to choose \epochs{} peptides from the datasets, the tasks of the datasets are disparate, and some datasets are small. Significant exceptions are the SHP-2 and soluble datasets, which show poor performance with limited examples relative to baseline models. In particular, SHP-2 seems to require the full dataset to achieve good accuracy. This may be due to the importance of motifs in this dataset,\cite{White2013b} and lack of feature re-use between datasets due to its unique task (binding affinity with specific enzyme).

The QBC training results are reported in Figure~\ref{fig:qbc-results}. QBC does not consistently provide better performance than random choice or uncertainty minimization. QBC significantly improves with meta-learning. As shown in Figure~\ref{fig:barchart}, QBC seems to have the best few-shot and final performance when combined with meta-learning, though uncertainty minimization is as good for few-shot only.  Supporting Information contains tabulated AUCs and accuracies in Tables S1-S8.

\begin{figure}
    \centering
    \includegraphics[width=0.8\linewidth]{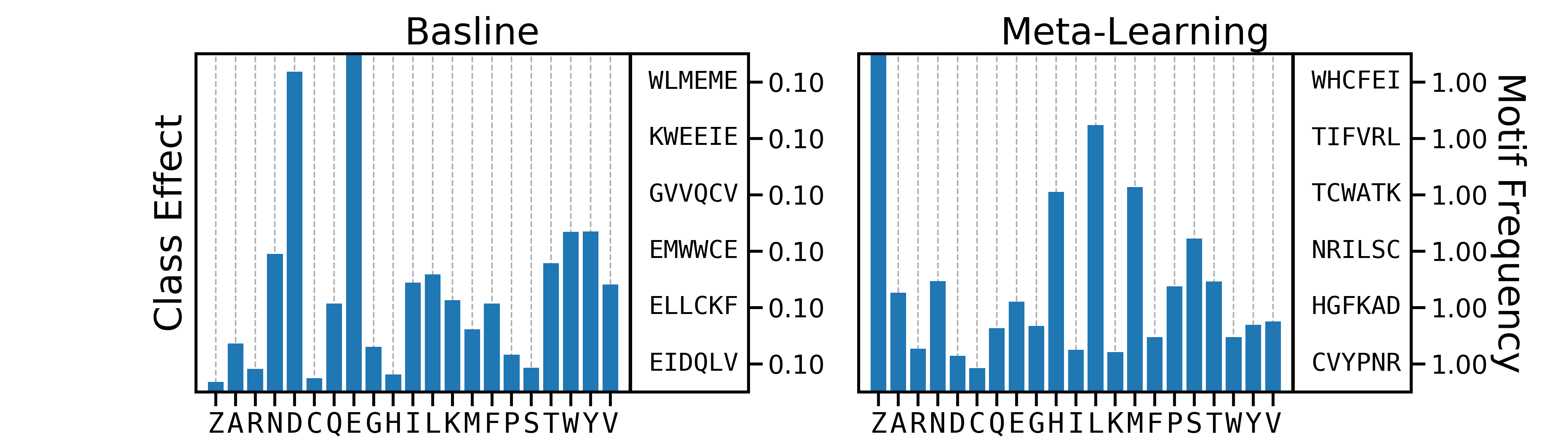}
    \caption{Features across tasks for baseline model. The barplot shows the partial derivative of probability of activity with respect to amino acid count averaged across training data. This gives importance for assigning label. The right side of the plots shows the maximum magnitude weight in the convolution, which roughly corresponds to most attended motif in the sequence. The y-axis label on the right side shows the frequency of this motif across \epochs{} training iterations. ``Z'' is the normalized length of the peptide.}
    \label{fig:interpret-all}
\end{figure}

Overall, meta-learning usually improves accuracy when combined with active learning methods across these datasets, especially , as shown in Figure~\ref{fig:barchart}. Recent analysis of meta-learning has shown mixed results across tasks. \citet{raghu2019rapid} showed that model agnostic meta-learning methods like Reptile only learn to re-use features across tasks. To assess how feature re-use can  be applicable in these dataset, we performed a basic sensitivity analysis in Figure~\ref{fig:interpret-all} which gives insight into the various features used in the modeling. Figure~\ref{fig:interpret-all} shows the features for the baseline model on the antibacterial dataset and the zero-shot features for meta-learning. Note that all motif frequencies are 1.0 since this is the meta-learned parameters, not a realization of training. The results show that meta-learning does not have the same features found in the baseline model, although some of the important amino acids are shared (N, C, H). Some of the amino acids within the motifs are shared, but they are not identical. 

\begin{figure}
    \centering
    \includegraphics[width=\linewidth]{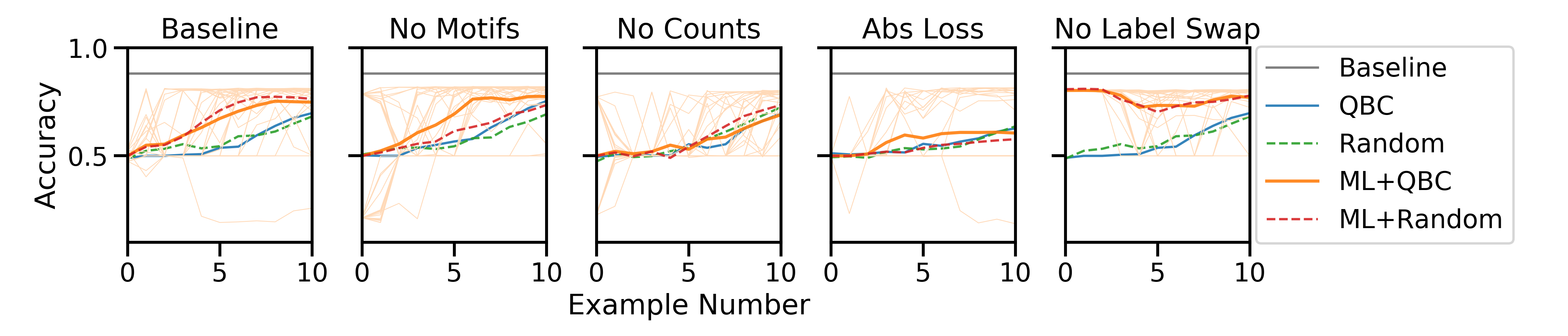}
    \caption{Training curves for different model choices, using the Antibacterial dataset. Baseline is the same in each panel and the baseline subplot is the model as presented in text. ``No Motifs'' has the convolution layers removed, ``No Counts'' has amino acid counts removed, ``Abs Loss'' uses an absolute difference loss instead of cross-entropy, and ``No Label Swap'' means that labels weren't swapped during meta-learning so that zero-shot accuracy is maximized. These results show that meta-learning is not always better, but is consistently a good choice for few-shot learning. }
    \label{fig:ablation}
\end{figure}

To ensure our conclusions about meta-learning and QBC being preferred for peptide design are robust, we explored four alternative model choices. These are shown in Figure~\ref{fig:ablation} for only the antibacterial task (although meta-learning traces were trained on all but antibacterial). The second subplot is ablation of the motif convolutions, i.e. using only amino acid count vectors for training. Without the motifs, accuracy is roughly the same, and only the combination of QBC with meta-learning is advantageous over random choice. Training with only motifs (convolutions) and no amino acid counts reduces performance in general and entirely removes the benefits of meta-learning, active learning, and the combination of the two. Using the original model structue but switching from a cross-entropy loss to absolute error loss reduces accuracy for all methods. The last plot shows what happens if we remove the label swapping during meta-learning, which is used to reduce over-fitting to ``active'' peptides. Zero-shot performance is improved, due to good correlation of activity in peptides across tasks. Meta-learning is preferred in this setting, when it is known that the class labels of active vs inactive can be re-used across tasks. These results indicate that the benefits of meta-learning paired with active learning can be sensitive to the model structure used. In particular, if the chosen loss function is poor, few-shot accuracy improvement may be impossible. The benefits of meta-learning combined with active learning seem to depend on the neural network features, as evidenced by the difference between the ``No Motifs'' and ``No Counts'' subplots of Figure~\ref{fig:ablation}.

Finally, to evaluate whether beta calibration\cite{Kull2017BetaCalibration} interacted favorably with these methods, the uncertainty minimization trials were repeated with beta calibration of the network output. In this case, beta calibration fitting is done separately after each training iteration, i.e. once per choice of new training point(s). The intent of beta calibration is to ensure that the scores output by the neural network reflect the actual probabilities that a given point belongs to the positive class. However, beta calibration did not have an appreciable effect on the accuracy or training efficiency (see Figure S2).

We assess statistical significance in Figure~\ref{fig:barchart}. The box-and-whisker plots show average accuracy values for the 10-example and 50-example training runs described in the Methods section. We performed Wilcoxon's signed ranks test on the 12 training accuracy values for a given method, comparing between all possible pairs of the six methods explored here, obtaining $p$-valuesfor each pair. Asterisks are shown in Figure~\ref{fig:barchart} for pairs with $p\leq 0.05$. Methods with no asterisk were not significantly different. Exact values for all $p$-values for accuracy can be found in the Supporting Information in Tables S7 and S8. We performed a similar analysis for the average AUC values for each method, also found in the Supporting Information in Figure S1 with $p$-values in Tables S5 and S6. We can see from Figure~\ref{fig:barchart} that after \epochs{} training examples, nearly all the methods explored here are not significantly different on average from random choice (with or without meta-learning). In the few-shot case, with only 10 training examples, the differences between methods are more pronounced. In this case, combining meta-learning with either uncertainty minimization or QBC yields significantly better average accuracy than either active learning method alone. This supports the hypothesis that meta-learning can enhance the accuracy of few-shot learning across multiple tasks, even for a relatively simple convolutional neural network.

\section{Conclusions}

This work has presented a dataset of 12 different peptide classification tasks, and explored active learning and meta-learning strategies for predicting peptide activity on them. The simple deep convolutional neural network used here offers greater than 85\% accuracy across all tasks except soluble, which is known to be a difficult task and was within reported ranges for other state-of-the-art models. We expect more complex models with attention, more layers, and long-range interactions in sequence space could improve the accuracy. The two active learning strategies explored here were not found to provide significant improvements over sampling peptides randomly.  Meta-learning was found to improve few-shot accuracy with 10 training examples only when combined with active learning. Here, every meta-learning method performed significantly better on average than uncertainty minimization with no meta-learning. Both meta-learning with uncertainty minimization and meta-learning with QBC were significantly better than either active learning srategy alone. In contrast, after \epochs{} training examples, only meta-learning with QBC was found to significantly increase average accuracy over random choice, indicating that the benefits of meta-learning are short-lived, but well-suited to a data-scarce context. These conclusions also hold in the zero-shot learning setting, but model ablation shows that they are dependent on loss choice and model features. This work provides a new peptide multi-task dataset and benchmark results for standard active learning and meta-learning methods, and shows that meta-learning could have significant benefits for experimental settings where data is scarce.

\section{Data Availability}
The datasets used in this work are available in the GitHub repository for the code: \url{https://github.com/ur-whitelab/peptide-ai}.

\section{Code Availability}
The code used to produce the results shown here is available in a GitHub repository: \url{https://github.com/ur-whitelab/peptide-ai}.

\section{Acknowledements}
This material is based upon work supported by the National Science Foundation under grants 1764415 and 1751471, and the Maximizing Investigators' Research Award Grant \#R35 GM137966 by the National Institute of General Medical Sciences under the National Institute of Health.

\bibliographystyle{biophysj}
\bibliography{bibliography.bib}

\end{document}


\maketitle

\begin{table}[ht]
    \centering
    \begin{tabular}{|c|ccccccc|}
        \hline
         & Baseline & Random & QBC & Umin & ML+Random & ML+QBC & ML+Umin\\\hline
Antibacterial & $ 0.94$ & 
$ 0.86 \pm 0.03$ & 
$ 0.86 \pm 0.10$ & 
$ 0.89 \pm 0.02$ & 
$ 0.87 \pm 0.05$ & 
$ 0.85 \pm 0.12$ & 
$ 0.85 \pm 0.13$  
\\
Anticancer & $ 0.86$ & 
$ 0.85 \pm 0.03$ & 
$ 0.79 \pm 0.18$ & 
$ 0.79 \pm 0.13$ & 
$ 0.87 \pm 0.03$ & 
$ 0.85 \pm 0.03$ & 
$ 0.84 \pm 0.06$  
\\
Antifungal & $ 0.92$ & 
$ 0.85 \pm 0.02$ & 
$ 0.83 \pm 0.07$ & 
$ 0.86 \pm 0.02$ & 
$ 0.84 \pm 0.03$ & 
$ 0.84 \pm 0.03$ & 
$ 0.87 \pm 0.02$  
\\
Anti-HIV & $ 0.86$ & 
$ 0.86 \pm 0.08$ & 
$ 0.75 \pm 0.12$ & 
$ 0.85 \pm 0.16$ & 
$ 0.78 \pm 0.05$ & 
$ 0.80 \pm 0.06$ & 
$ 0.87 \pm 0.04$  
\\
Anti-MRSA & $ 1.00$ & 
$ 0.95 \pm 0.02$ & 
$ 0.95 \pm 0.03$ & 
$ 0.93 \pm 0.19$ & 
$ 0.96 \pm 0.03$ & 
$ 0.96 \pm 0.03$ & 
$ 0.94 \pm 0.05$  
\\
Antiparasital & $ 0.89$ & 
$ 0.95 \pm 0.05$ & 
$ 0.93 \pm 0.13$ & 
$ 0.94 \pm 0.16$ & 
$ 0.93 \pm 0.03$ & 
$ 0.97 \pm 0.02$ & 
$ 0.96 \pm 0.02$  
\\
Antiviral & $ 0.85$ & 
$ 0.83 \pm 0.05$ & 
$ 0.78 \pm 0.10$ & 
$ 0.81 \pm 0.19$ & 
$ 0.74 \pm 0.05$ & 
$ 0.87 \pm 0.05$ & 
$ 0.81 \pm 0.14$  
\\
Hemolytic & $ 0.99$ & 
$ 0.99 \pm 0.00$ & 
$ 0.92 \pm 0.21$ & 
$ 0.88 \pm 0.24$ & 
$ 0.98 \pm 0.01$ & 
$ 0.97 \pm 0.06$ & 
$ 0.93 \pm 0.14$  
\\
Soluble & $ 0.63$ & 
$ 0.53 \pm 0.02$ & 
$ 0.53 \pm 0.03$ & 
$ 0.53 \pm 0.02$ & 
$ 0.53 \pm 0.02$ & 
$ 0.53 \pm 0.03$ & 
$ 0.52 \pm 0.02$  
\\
SHP-2 & $ 0.97$ & 
$ 0.87 \pm 0.04$ & 
$ 0.87 \pm 0.04$ & 
$ 0.89 \pm 0.02$ & 
$ 0.91 \pm 0.05$ & 
$ 0.90 \pm 0.03$ & 
$ 0.86 \pm 0.04$  
\\
TULA-2 & $ 0.90$ & 
$ 0.86 \pm 0.05$ & 
$ 0.91 \pm 0.03$ & 
$ 0.88 \pm 0.04$ & 
$ 0.95 \pm 0.02$ & 
$ 0.91 \pm 0.04$ & 
$ 0.95 \pm 0.03$  
\\
Human & $ 0.94$ & 
$ 0.89 \pm 0.01$ & 
$ 0.84 \pm 0.19$ & 
$ 0.82 \pm 0.20$ & 
$ 0.89 \pm 0.02$ & 
$ 0.90 \pm 0.02$ & 
$ 0.89 \pm 0.01$  
\\
\hline
    \end{tabular}
    \vspace{0.2cm}

    \caption{Area under curve (AUC) for receiver operator characteristic curves for classifiers trained on 50 datapoints with different active learning methods on different datasets. Baseline was trained on all data, whereas others saw 50 peptides according to their active learning strategy. Errors are computed from standard deviation across 100 trials on different data splits and random sampling in active learning strategy. Umin is uncertainty minimization, ML is meta-learning, and QBC is query by committee.}
    \label{tab:auc_50}
\end{table}

\begin{table}[ht]
    \centering
    \begin{tabular}{|c|ccccccc|}
        \hline
         & Baseline & Random & QBC & Umin & ML+Random & ML+QBC & ML+Umin\\\hline
Antibacterial & $ 0.94$ & 
$ 0.84 \pm 0.01$ & 
$ 0.83 \pm 0.01$ & 
$ 0.74 \pm 0.21$ & 
$ 0.84 \pm 0.01$ & 
$ 0.80 \pm 0.11$ & 
$ 0.83 \pm 0.02$  
\\
Anticancer & $ 0.86$ & 
$ 0.87 \pm 0.15$ & 
$ 0.88 \pm 0.01$ & 
$ 0.77 \pm 0.19$ & 
$ 0.80 \pm 0.03$ & 
$ 0.84 \pm 0.05$ & 
$ 0.90 \pm 0.02$  
\\
Antifungal & $ 0.92$ & 
$ 0.81 \pm 0.12$ & 
$ 0.79 \pm 0.20$ & 
$ 0.73 \pm 0.24$ & 
$ 0.84 \pm 0.02$ & 
$ 0.82 \pm 0.02$ & 
$ 0.75 \pm 0.24$  
\\
Anti-HIV & $ 0.86$ & 
$ 0.89 \pm 0.14$ & 
$ 0.76 \pm 0.22$ & 
$ 0.79 \pm 0.28$ & 
$ 0.86 \pm 0.13$ & 
$ 0.79 \pm 0.15$ & 
$ 0.85 \pm 0.13$  
\\
Anti-MRSA & $ 1.00$ & 
$ 1.00 \pm 0.01$ & 
$ 0.96 \pm 0.17$ & 
$ 0.84 \pm 0.30$ & 
$ 0.92 \pm 0.01$ & 
$ 0.95 \pm 0.01$ & 
$ 0.95 \pm 0.01$  
\\
Antiparasital & $ 0.89$ & 
$ 0.94 \pm 0.17$ & 
$ 0.93 \pm 0.25$ & 
$ 0.91 \pm 0.24$ & 
$ 0.90 \pm 0.02$ & 
$ 0.95 \pm 0.02$ & 
$ 0.95 \pm 0.01$  
\\
Antiviral & $ 0.85$ & 
$ 0.88 \pm 0.09$ & 
$ 0.75 \pm 0.14$ & 
$ 0.78 \pm 0.20$ & 
$ 0.86 \pm 0.14$ & 
$ 0.84 \pm 0.13$ & 
$ 0.82 \pm 0.13$  
\\
Hemolytic & $ 0.99$ & 
$ 0.94 \pm 0.05$ & 
$ 0.82 \pm 0.24$ & 
$ 0.89 \pm 0.21$ & 
$ 0.81 \pm 0.10$ & 
$ 0.76 \pm 0.18$ & 
$ 0.85 \pm 0.12$  
\\
Soluble & $ 0.63$ & 
$ 0.51 \pm 0.03$ & 
$ 0.50 \pm 0.02$ & 
$ 0.50 \pm 0.03$ & 
$ 0.51 \pm 0.02$ & 
$ 0.51 \pm 0.02$ & 
$ 0.51 \pm 0.03$  
\\
SHP-2 & $ 0.97$ & 
$ 0.66 \pm 0.13$ & 
$ 0.77 \pm 0.08$ & 
$ 0.69 \pm 0.11$ & 
$ 0.61 \pm 0.11$ & 
$ 0.69 \pm 0.17$ & 
$ 0.58 \pm 0.22$  
\\
TULA-2 & $ 0.90$ & 
$ 0.71 \pm 0.09$ & 
$ 0.80 \pm 0.09$ & 
$ 0.79 \pm 0.14$ & 
$ 0.60 \pm 0.16$ & 
$ 0.72 \pm 0.15$ & 
$ 0.66 \pm 0.16$  
\\
Human & $ 0.94$ & 
$ 0.84 \pm 0.13$ & 
$ 0.83 \pm 0.18$ & 
$ 0.80 \pm 0.23$ & 
$ 0.87 \pm 0.01$ & 
$ 0.85 \pm 0.00$ & 
$ 0.87 \pm 0.01$  
\\
\hline
    \end{tabular}
    \vspace{0.2cm}

    \caption{Area under curve (AUC) for receiver operator characteristic curves for classifiers trained on 10 datapoints with different active learning methods on different datasets. Baseline was trained on all data, whereas others saw 10 peptides according to their active learning strategy. Errors are computed from standard deviation across 30 trials on different data splits and random sampling in active learning strategy. Umin is uncertainty minimization, ML is meta-learning, and QBC is query by committee.}
    \label{tab:auc_10}
\end{table}

\begin{table}[ht]
    \centering
    \begin{tabular}{|c|ccccccc|}
        \hline
         & Baseline & Random & QBC & Umin & ML+Random & ML+QBC & ML+Umin\\\hline

Antibacterial & $ 0.88$ & 
$ 0.78 \pm 0.04$ & 
$ 0.78 \pm 0.08$ & 
$ 0.81 \pm 0.02$ & 
$ 0.80 \pm 0.04$ & 
$ 0.80 \pm 0.03$ & 
$ 0.79 \pm 0.09$  
\\
Anticancer & $ 0.82$ & 
$ 0.79 \pm 0.04$ & 
$ 0.78 \pm 0.07$ & 
$ 0.76 \pm 0.08$ & 
$ 0.80 \pm 0.05$ & 
$ 0.82 \pm 0.02$ & 
$ 0.78 \pm 0.06$  
\\
Antifungal & $ 0.87$ & 
$ 0.77 \pm 0.05$ & 
$ 0.78 \pm 0.04$ & 
$ 0.81 \pm 0.02$ & 
$ 0.76 \pm 0.04$ & 
$ 0.80 \pm 0.03$ & 
$ 0.82 \pm 0.03$  
\\
Anti-HIV & $ 0.84$ & 
$ 0.80 \pm 0.08$ & 
$ 0.79 \pm 0.05$ & 
$ 0.79 \pm 0.09$ & 
$ 0.71 \pm 0.05$ & 
$ 0.85 \pm 0.05$ & 
$ 0.82 \pm 0.04$  
\\
Anti-MRSA & $ 0.96$ & 
$ 0.90 \pm 0.03$ & 
$ 0.95 \pm 0.02$ & 
$ 0.94 \pm 0.12$ & 
$ 0.95 \pm 0.04$ & 
$ 0.95 \pm 0.02$ & 
$ 0.93 \pm 0.01$  
\\
Antiparasital & $ 0.89$ & 
$ 0.89 \pm 0.07$ & 
$ 0.91 \pm 0.08$ & 
$ 0.91 \pm 0.11$ & 
$ 0.90 \pm 0.03$ & 
$ 0.90 \pm 0.03$ & 
$ 0.94 \pm 0.03$  
\\
Antiviral & $ 0.76$ & 
$ 0.75 \pm 0.06$ & 
$ 0.74 \pm 0.04$ & 
$ 0.80 \pm 0.11$ & 
$ 0.69 \pm 0.05$ & 
$ 0.77 \pm 0.04$ & 
$ 0.76 \pm 0.08$  
\\
Hemolytic & $ 0.98$ & 
$ 0.98 \pm 0.01$ & 
$ 0.91 \pm 0.12$ & 
$ 0.90 \pm 0.12$ & 
$ 0.95 \pm 0.01$ & 
$ 0.96 \pm 0.03$ & 
$ 0.94 \pm 0.07$  
\\
Soluble & $ 0.59$ & 
$ 0.52 \pm 0.02$ & 
$ 0.51 \pm 0.01$ & 
$ 0.52 \pm 0.02$ & 
$ 0.52 \pm 0.02$ & 
$ 0.52 \pm 0.02$ & 
$ 0.51 \pm 0.02$  
\\
SHP-2 & $ 0.90$ & 
$ 0.78 \pm 0.04$ & 
$ 0.82 \pm 0.05$ & 
$ 0.80 \pm 0.04$ & 
$ 0.77 \pm 0.11$ & 
$ 0.79 \pm 0.09$ & 
$ 0.77 \pm 0.07$  
\\
TULA-2 & $ 0.83$ & 
$ 0.77 \pm 0.06$ & 
$ 0.81 \pm 0.05$ & 
$ 0.81 \pm 0.05$ & 
$ 0.85 \pm 0.10$ & 
$ 0.82 \pm 0.05$ & 
$ 0.85 \pm 0.05$  
\\
Human & $ 0.89$ & 
$ 0.82 \pm 0.03$ & 
$ 0.85 \pm 0.07$ & 
$ 0.80 \pm 0.12$ & 
$ 0.83 \pm 0.03$ & 
$ 0.84 \pm 0.03$ & 
$ 0.84 \pm 0.03$  
\\
\hline
    \end{tabular}
    \vspace{0.2cm}

    \caption{Final training accuracy values for classifiers trained with different active learning methods on different datasets. Baseline was trained on all data, whereas others saw 50 peptides according to their active learning strategy. Errors are computed from standard deviation across 30 trials on different data splits and random sampling in active learning strategy. Umin is uncertainty minimization, ML is meta-learning, and QBC is query by committee.}
    \label{tab:acc_50}    
\end{table}

\begin{table}[ht]
    \centering
    \begin{tabular}{|c|ccccccc|}
        \hline
         & Baseline & Random & QBC & Umin & ML+Random & ML+QBC & ML+Umin\\\hline

antibacterial & $ 0.88$ & 
$ 0.74 \pm 0.10$ & 
$ 0.75 \pm 0.06$ & 
$ 0.66 \pm 0.13$ & 
$ 0.79 \pm 0.03$ & 
$ 0.73 \pm 0.10$ & 
$ 0.79 \pm 0.06$  
\\
anticancer & $ 0.82$ & 
$ 0.77 \pm 0.13$ & 
$ 0.79 \pm 0.10$ & 
$ 0.71 \pm 0.11$ & 
$ 0.76 \pm 0.08$ & 
$ 0.81 \pm 0.07$ & 
$ 0.86 \pm 0.02$  
\\
antifungal & $ 0.87$ & 
$ 0.71 \pm 0.12$ & 
$ 0.71 \pm 0.14$ & 
$ 0.67 \pm 0.14$ & 
$ 0.79 \pm 0.06$ & 
$ 0.78 \pm 0.06$ & 
$ 0.74 \pm 0.14$  
\\
anti-HIV & $ 0.84$ & 
$ 0.71 \pm 0.19$ & 
$ 0.73 \pm 0.15$ & 
$ 0.75 \pm 0.15$ & 
$ 0.79 \pm 0.11$ & 
$ 0.77 \pm 0.10$ & 
$ 0.79 \pm 0.13$  
\\
anti-MRSA & $ 0.96$ & 
$ 0.93 \pm 0.13$ & 
$ 0.92 \pm 0.12$ & 
$ 0.82 \pm 0.15$ & 
$ 0.87 \pm 0.07$ & 
$ 0.95 \pm 0.01$ & 
$ 0.93 \pm 0.02$  
\\
antiparasital & $ 0.89$ & 
$ 0.86 \pm 0.15$ & 
$ 0.81 \pm 0.16$ & 
$ 0.83 \pm 0.15$ & 
$ 0.85 \pm 0.11$ & 
$ 0.92 \pm 0.08$ & 
$ 0.92 \pm 0.02$  
\\
antiviral & $ 0.76$ & 
$ 0.73 \pm 0.15$ & 
$ 0.64 \pm 0.11$ & 
$ 0.71 \pm 0.10$ & 
$ 0.77 \pm 0.09$ & 
$ 0.79 \pm 0.08$ & 
$ 0.77 \pm 0.09$  
\\
hemolytic & $ 0.98$ & 
$ 0.79 \pm 0.11$ & 
$ 0.70 \pm 0.15$ & 
$ 0.79 \pm 0.13$ & 
$ 0.79 \pm 0.09$ & 
$ 0.75 \pm 0.11$ & 
$ 0.85 \pm 0.02$  
\\
soluble & $ 0.59$ & 
$ 0.50 \pm 0.01$ & 
$ 0.50 \pm 0.01$ & 
$ 0.50 \pm 0.01$ & 
$ 0.50 \pm 0.01$ & 
$ 0.51 \pm 0.01$ & 
$ 0.50 \pm 0.02$  
\\
SHP-2 & $ 0.90$ & 
$ 0.51 \pm 0.02$ & 
$ 0.51 \pm 0.03$ & 
$ 0.51 \pm 0.03$ & 
$ 0.50 \pm 0.00$ & 
$ 0.50 \pm 0.00$ & 
$ 0.51 \pm 0.04$  
\\
TULA-2 & $ 0.83$ & 
$ 0.51 \pm 0.04$ & 
$ 0.52 \pm 0.05$ & 
$ 0.51 \pm 0.05$ & 
$ 0.50 \pm 0.00$ & 
$ 0.50 \pm 0.00$ & 
$ 0.50 \pm 0.00$  
\\
human & $ 0.89$ & 
$ 0.83 \pm 0.07$ & 
$ 0.80 \pm 0.12$ & 
$ 0.77 \pm 0.16$ & 
$ 0.83 \pm 0.09$ & 
$ 0.81 \pm 0.09$ & 
$ 0.83 \pm 0.09$  
\\
\hline
    \end{tabular}
    \vspace{0.2cm}

    \caption{Final training accuracy values for classifiers trained with different active learning methods on different datasets. Baseline was trained on all data, whereas others saw 10 peptides according to their active learning strategy. Errors are computed from standard deviation across 100 trials on different data splits and random sampling in active learning strategy. Umin is uncertainty minimization, ML is meta-learning, and QBC is query by committee.}
    \label{tab:acc_10}    
\end{table}

\begin{table}
    \centering
    \begin{tabular}{|c|cccccc|}
        \hline & Random & Umin & QBC & ml+random & ML+Umin & ML+QBC\\\hline
        Random & N/A & 0.5303 & 0.7537 & 0.8139 & 0.8753 & \textbf{0.04139 }\\
        Umin & 0.5303 & N/A & 0.3078 & 0.3465 & 0.1361 & \textbf{0.0186 }\\
        QBC & 0.7537 & 0.3078 & N/A & 0.4802 & 0.3078 & \textbf{0.01206 }\\
        ML+Random & 0.8139 & 0.3465 & 0.4802 & N/A & 0.9375 & 0.2094 \\
        ML+Umin & 0.8753 & 0.1361 & 0.3078 & 0.9375 & N/A & 0.2094 \\
        ML+QBC & \textbf{0.04139 }& \textbf{0.0186 }& \textbf{0.01206 }& 0.2094 & 0.2094 & N/A \\\hline
    \end{tabular}
    \caption{Area under curve (AUC) $p$-values between different methods after training on 50 peptides. These $p$-values were calculated using Wilcoxon's signed ranks test among the twelve average AUC values given by the corresponding pair of methods. Bold indicates significance ($p \leq 0.05$).}
    \label{tab:auc_pvals_50}
\end{table}

\begin{table}
    \centering
    \begin{tabular}{|c|cccccc|}
        \hline & Random & Umin & QBC & ml+random & ML+Umin & ML+QBC\\\hline
        Random & N/A & \textbf{0.01502 }& 0.1167 & \textbf{0.02806 }& \textbf{0.03417 }& 0.1823 \\
        Umin & \textbf{0.01502 }& N/A & 0.1167 & 0.5303 & 0.2721 & 0.1579 \\
        QBC & 0.1167 & 0.1167 & N/A & 0.8139 & 0.6379 & 0.6379 \\
        ML+Random & \textbf{0.02806 }& 0.5303 & 0.8139 & N/A & 0.5829 & 0.8139 \\
        ML+Umin & \textbf{0.03417 }& 0.2721 & 0.6379 & 0.5829 & N/A & 0.8139 \\
        ML+QBC & 0.1823 & 0.1579 & 0.6379 & 0.8139 & 0.8139 & N/A \\\hline
    \end{tabular}
    \caption{Area under curve (AUC) $p$-values between different methods after training on 10 peptides. These $p$-values were calculated using Wilcoxon's signed ranks test among the twelve average AUC values given by the corresponding pair of methods. Bold indicates significance ($p \leq 0.05$).}
    \label{tab:auc_pvals_10}
\end{table}

\begin{table}
    \centering
    \begin{tabular}{|c|cccccc|}
        \hline & Random & Umin & QBC & ml+random & ML+Umin & ML+QBC\\\hline
        Random & N/A & 0.3078 & 0.4802 & 1.0 & 0.1167 & \textbf{0.007649 }\\
        Umin & 0.3078 & N/A & 1.0 & 0.6379 & 0.3078 & 0.2721 \\
        QBC & 0.4802 & 1.0 & N/A & 0.5303 & 0.2094 & \textbf{0.04986 }\\
        ML+Random & 1.0 & 0.6379 & 0.5303 & N/A & 0.3078 & \textbf{0.02806 }\\
        ML+Umin & 0.1167 & 0.3078 & 0.2094 & 0.3078 & N/A & 0.3882 \\
        ML+QBC & \textbf{0.007649 }& 0.2721 & \textbf{0.04986 }& \textbf{0.02806 }& 0.3882 & N/A \\\hline
    \end{tabular}
    \caption{Final training accuracy $p$-values between different methods after training on 50 peptides. These $p$-values were calculated using Wilcoxon's signed ranks test among the twelve average accuracy values given by the corresponding pair of methods. Bold indicates significance ($p \leq 0.05$).}
    \label{tab:acc_pvals_50}
\end{table}

\begin{table}
    \centering
    \begin{tabular}{|c|cccccc|}
        \hline & Random & Umin & QBC & ml+random & ML+Umin & ML+QBC\\\hline
        Random & N/A & 0.07119 & 0.4328 & 0.8753 & \textbf{0.0186 }& 0.1579 \\
        Umin & 0.07119 & N/A & 0.4802 & \textbf{0.02291 }& \textbf{0.007646 }& \textbf{0.02806 }\\
        QBC & 0.4328 & 0.4802 & N/A & 0.1167 & \textbf{0.009633 }& \textbf{0.03417 }\\
        ML+Random & 0.8753 & \textbf{0.02291 }& 0.1167 & N/A & 0.05046 & 0.7989 \\
        ML+Umin & \textbf{0.0186 }& \textbf{0.007646 }& \textbf{0.009633 }& 0.05046 & N/A & 0.2477 \\
        ML+QBC & 0.1579 & \textbf{0.02806 }& \textbf{0.03417 }& 0.7989 & 0.2477 & N/A \\\hline
    \end{tabular}
    \caption{Final training accuracy $p$-values between different methods after training on 10 peptides. These $p$-values were calculated using Wilcoxon's signed ranks test among the twelve average accuracy values given by the corresponding pair of methods. Bold indicates significance ($p \leq 0.05$).}
    \label{tab:acc_pvals_10}
\end{table}

\begin{figure}
    \centering
    \includegraphics[width=\textwidth]{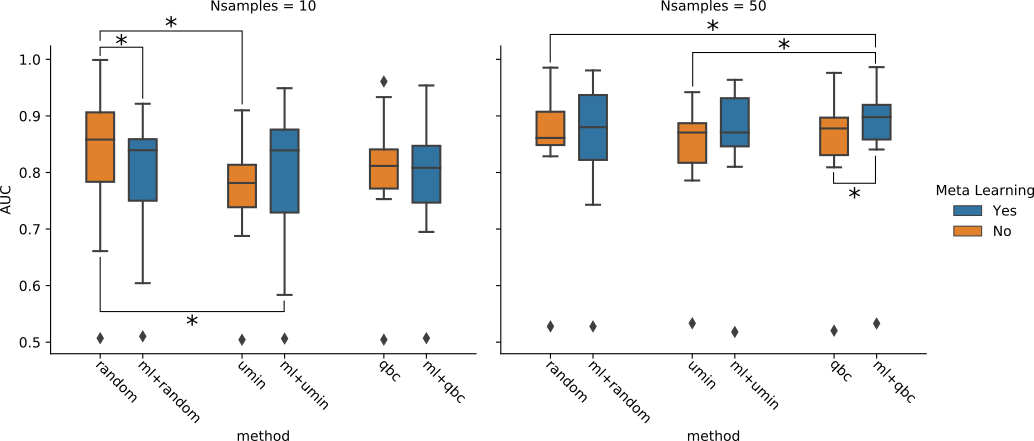}
    \caption{Box-and-whisker plot comparing average AUC values after 10 and 50 training examples across all twelve datasets for five different methods explored in this work. Asterisks indicate statistically significant difference ($p\leq 0.05$) in mean AUC, calculated using Wilcoxon's signed ranks test among the twelve average AUC values compared between two methods.}
    \label{fig:barchart}
\end{figure}

\begin{figure}
    \centering
    \includegraphics[width=\textwidth]{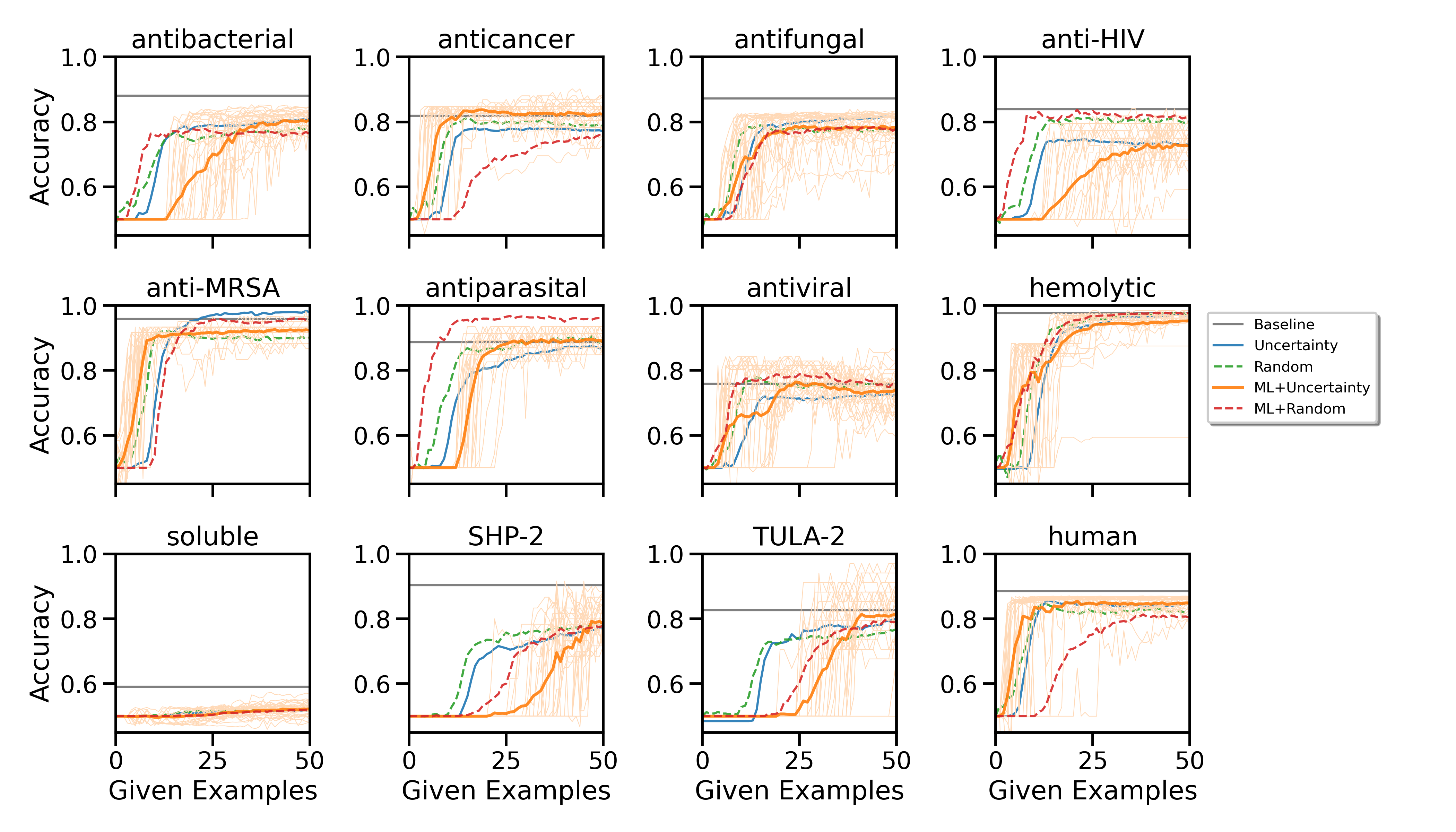}
    \caption{Training curves of uncertainty minimization active learning with beta calibration compared with baseline (gray) trained across all data points and randomly choosing examples. The y-axis is accuracy on withheld data. Light traces are individual 30 runs and the dark trace is median (only one set of traces is shown). Each run has a different train/withheld split and random number generator seeds. Each subplot is a different task.}
    \label{fig:training}
\end{figure}